\documentclass{article}
\usepackage{arxiv}
\usepackage{amsmath,amssymb,amsfonts}
\usepackage{algorithmic}
\usepackage{algorithm}
\usepackage{array}
\usepackage[caption=false,font=normalsize,labelfont=sf,textfont=sf]{subfig}
\usepackage{textcomp}
\usepackage{stfloats}
\usepackage{url}
\usepackage{verbatim}
\usepackage{graphicx}
\usepackage[table]{xcolor}
\usepackage{tabularx}
\usepackage{enumitem}
\usepackage{ulem}
\usepackage{lmodern,textcomp}
\usepackage[numbers]{natbib} 
\usepackage{multirow}
\usepackage{framed}
\usepackage{booktabs}
\usepackage{tikz}
\usepackage{color,soul}
\usepackage{adjustbox}
\usepackage{threeparttable}
\usepackage{orcidlink}
\def\BibTeX{{\rm B\kern-.05em{\sc i\kern-.025em b}\kern-.08em
    T\kern-.1667em\lower.7ex\hbox{E}\kern-.125emX}}

\title{Beyond Literacy: Predicting Interpretation Correctness of Visualizations with User Traits, Item Difficulty, and Rasch Scores}

\author{Davide Falessi\thanks{Corresponding author.} \\
	Department of Informatics Engineering\\
	Universit{\`a} degli Studi di Roma Tor Vergata\\
	Rome, Italy\\
	\texttt{falessi@ing.uniroma2.it} \\
	\And
	Silvia Golia \\
	Department of Economics and Management\\
	Universit{\`a} degli Studi di Brescia\\
	Brescia, Italy\\
	\texttt{silvia.golia@unibs.it} \\
	\AND
	Agela Locoro \\
	Department of Economics and Management\\
	Universit{\`a} degli Studi di Brescia\\
	Brescia, Italy\\
	\texttt{angela.locoro@unibs.it} \\
	Manuel Mastrofini \\
	Department of Informatics Engineering\\
	Universit{\`a} degli Studi di Roma Tor Vergata\\
	Rome, Italy\\
	\texttt{mastrofini@ing.uniroma2.it} \\
}

\renewcommand{\shorttitle}{\textit{arXiv} Template}

\hypersetup{
pdftitle={A template for the arxiv style},
pdfauthor={Davide~Falessi, Silvia~Golia, Angela~Locoro, Manuel~Mastrofini},
pdfkeywords={Visualization Literacy, Machine Learning, Human Performance Prediction,
Human-Data Interaction, Feature Importance, Adaptive Assessment Design},
}

\begin{document}
\maketitle

\begin{abstract}
Data Visualization Literacy assessments are typically administered via fixed sets of Data Visualization items, despite substantial heterogeneity in how different people interpret the same visualization.  This paper presents and evaluates an approach for predicting Human Interpretation Correctness (P-HIC) of data visualizations; i.e., anticipating whether a specific person will interpret a data visualization correctly or not, before exposure to that DV, enabling more personalized assessment and training.
We operationalize P-HIC as a binary classification problem using 22 features spanning Human Profile, Human Performance, and Item difficulty (including ExpertDifficulty and RaschDifficulty). We evaluate three machine-learning models (Logistic Regression model, Random Forest, Multi Layer Perceptron) with and without feature selection, using a survey with 1,083 participants who answered 32 Data Visualization items (eight data visualizations per four items), yielding 34,656 item responses. Performance is assessed via a ten-time ten-fold cross-validation in each 32 (item-specific) datasets, using AUC and Cohen's kappa.
Logistic Regression model with feature selection is the best-performing approach, reaching a median AUC of 0.72 and a median kappa of 0.32. Feature analyses show RaschDifficulty as the dominant predictor, followed by experts’ ratings and prior correctness (PercCorrect), whose relevance increases across sessions. Profile information did not particularly support P-HIC.
Our results support the feasibility of anticipating misinterpretations of data visualizations, and motivate the runtime selection of data visualizations items tailored to an audience, thereby improving the efficiency of Data Visualization Literacy assessment and targeted training.
\end{abstract}


\section{Introduction}
\label{sec:intro}
\label{sec:Introduction}
\newcommand{\rqone}{RQ1: How does human interpretation correctness vary?}
\newcommand{\rqtwo}{RQ2: How accurately can we predict human interpretation correctness before interaction?}
\newcommand{\rqthree}{RQ3: Which prediction features are most informative of human interpretation correctness?}

Data Visualization Literacy (\textbf{DVL}) stands for the ability of adults to properly process information related to data visualization (\textbf{DV}) artifacts, i.e., encoding information into and decoding information from them~\cite{locoro2021visual,beschi2025characterizing}.

Recent studies have shown that more intelligent interfaces for supporting decision-making are heavily demanded, capable of supporting personalized interactions informed by users' profiles and characteristics, beyond basic usability~\cite{locoro2025modeling,beschi2025ai}. 

Visualization Literacy assessment tests exist to measure the level of skills required to correctly interpret DV,e.g., the VLAT~\cite{Lee2017551,Pandey20231} and mini-VLAT tests, which provide a longer and shorter version of the same assessment test, respectively. However, assessment test items are usually fixed in number, static, mono-tasks, and the only variability provided during their administration to users is the randomization order of items. Visualization design needs to define, model, and assess human ability and data visualization usability in a more personalized way, for better usability and decision-making. 
Since one size does not fit all, it is essential to identify which visualization will lead a specific person towards a correct interpretation and which is too hard to be effective. Therefore, there is a major need to personalize literacy assessment tests to make them more effective for real-world scenarios, where the goal is to achieve the best possible interpretation of data for each user in the shortest time possible, thereby reducing cognitive burden and lowering training costs.



%
This paper addresses the following three research questions.

\textit{\rqone}  Recent works have explored human performance on DV items using the following approaches: item difficulty, visual attention patterns, adaptive assessment, or the design of complementary empirical datasets comparing human and model responses~\cite{beschi2025characterizing}. Our approach fills the gap in integrating both qualitative and quantitative aspects of human performance by studying the characteristics of features that impact DVL and their predictive quality. Our contribution aims to go beyond mere literacy, i.e., human ability derived from its level of expertise or depending on items difficulty, towards a more comprehensive approach that integrates static and dynamic information. We define an \textbf{item} as a question on a DV, e.g., the highest point in or the name of the DV. Each item has one correct and many incorrect answers. Thus, we define human interpretation correctness (\textbf{HIC}) as a boolean status of a human answer to an item: yes, if the human's answer coincides with the correct answer, or no, otherwise. In this research question, we investigate the level of HIC and how this varies with question type and fatigue.

\textit{\rqtwo} Indeed, the same data can be visualized in multiple ways, such as a table, a bar chart, or a box plot, which are all admissible DV \cite{page2021prisma}. However, the interpretation of a DV by a human can vary depending on several factors, including the complexity of the DV, the individual's level of DVL, the estimated difficulty of the DV as judged by experts, the task at hand, and the human's contingent performance due to fatigue bias or overconfidence. In real case scenarios, misunderstandings of DV can lead to poor decision-making; hence, it is essential to anticipate whether a particular DV is likely to be correctly interpreted by a specific user. In this study, we propose and evaluate the first approach to predict HIC (\textbf{P-HIC}) before the human interacts, i.e., sees, the item.  In order to identify the best predictors for our problem, we use some well-established machine learning (\textbf{ML}) models, such as random forests, logistic regression model, and multi layer perceptron. P-HIC considers both quantitative (e.g., item difficulty score, experts' score, number of correct answers before the current answer) and qualitative aspects (e.g., users' ability, background knowledge), i.e., everything that can be measured before the human is exposed to the item. Furthermore, our modelling approach accounts for common emerging aspects of human performance, such as fatigue. We use the term \textbf{accuracy} to refer to the ability of P-HIC to predict HIC, not to interpret items.
     
\textit{\rqthree} P-HIC can leverage and combine the different characteristics of both humans and items. Understanding which feature or mix of features is important, i.e., supports P-HIC,  provides two benefits: I) it identifies which feature needs to be measured for supporting P-HIC-based approaches, and II) whenever some features result in actionable information (e.g., education degree), these could be boosted to improve visualization design and people's DVL \cite{Lee2017551,cui2023adaptive, beschi2025characterizing}.

In this work, we carried out a survey with 1,083 participants, who answered four questions per eight DVs, thus leading to 32 items and 34,656 answers. A ten-time ten-fold validation approach shows that Logistic Regression model with feature selection is the best performing model, reaching a mediam accuracy across humans and questions of 0.72 AUC and a Kappa of 0.32. Finally, the most important features resulted experts' rating, Rasch scores, and percentage of previous correct responses.

The remainder of this paper is structured as follows. Section \ref{sec:related} discusses the related literature, focusing in particular on empirical studies about visualization literacy, with a focus on users' profiling and assessment tests design. Section \ref{sec:design} reports the approach design, Section \ref{sec:results} shows the results, and Section \ref{sec:discussion} reports a discussion of our study implications and the threats to validity. Finally, Section \ref{sec:conclusions} concludes the paper and outlines directions for future work.

\section{Related Work}
\label{sec:related}
The work by~\citet{Cui-Promises-Pitfalls} grounded item design in established visualization task taxonomies and evaluated the resulting items through expert review and human performance studies. The resulting VILA-VLAT test was validated by measuring human performance across visualization questions and by demonstrating convergent validity with an existing visualization literacy assessment. This situates the paper alongside prior work that treats correctness as an outcome of both visual encoding properties (e.g., chart type, aggregation) and cognitive task demands, rather than as a purely individual trait.
Important for prediction of interpretation correctness research, the paper revealed that item characteristics—such as data aggregation level, visual density, and perceptual ambiguity—systematically affect whether humans answer correctly. Several of the pitfalls identified by the authors (e.g., misleading aggregation, ambiguous extrema, or visually underspecified comparisons) correspond to known sources of human error in visualization interpretation, reinforcing the idea that correctness is jointly determined by item structure and perceptual-cognitive constraints.
From the perspective of correctness prediction, this work can thus be seen as complementary: it expands the space of visualization items over which human correctness can be observed and analyzed, while underscoring the necessity of item modelling when interpreting human performance on visualization tasks. Our work expands it by underscoring the necessity of item selection and personalization.

Some works directly modelled variation in DV item difficulty based on human performance~\cite{verma2025measuring}, using multiple DVL tests to estimate item-level difficulty. In this context, task characteristics may explain performance variation. While not a full PIC model, this work was among the first to systematically quantify how item features may explain human correctness variability. Its relevancy to the present study relies on its operationalization of the core problem of explaining/predicting which items humans will answer correctly. Our work extends it by trying to outline how the characteristics of both subjects and items impact on the correctness of answers.

The work by~\citet{chang2025tell} used eye-tracking data across visualization tests and showed that distinct attention patterns correlate with correctness performance. Although the outcome was literacy level, not item answer correctness directly, literacy level was strongly tied to correctness and was predicted from observable behavior. This points toward eye-tracking features may be useful in correctness prediction. However, the granularity of features for these kinds of approaches is too fine. Our approach leverages more qualitative data, rather than visual perception behaviour.

\citet{verma2024evaluating} compared human and ML model performance on visualization literacy assessments.
Human correctness patterns were measured across different cohorts and task types. Empirical data on which visualization items humans answer correctly vs incorrectly, could feed predictive modeling and feature analysis.

Regarding adaptive tests, \citet{cui2023adaptive} developed adaptive tests (e.g., A-VLAT, A-CALVI) using item response theory (IRT), by explicitly modelling the probability a person with a given ability answered an item correctly based on item difficulty and discrimination parameters. However, this prediction model did not rely on ML. Our work tries to extend it to more complex predictors with DV task-based items.

Finally, works such as \citet{halkiopoulosGkintoni2024} supports the idea that adaptive and personalized predictive models can use user performance history to improve future predictions, aligning well with the idea that performance history gains importance over sessions. However, this research comes from the adaptive learning literature, and do not explicitly address the domain and issues of the DV research field. Our work creates a strong connection between these two fields, arguing that adaptive learning is better at mirroring the current fast evolving scenario in human learning and understanding.

\section{Method}
\label{sec:design}
This work is survey-centric, meaning it utilizes a survey to address all three research questions. Therefore, we structure this section by first presenting the survey and then the details on how we investigated each research question.

\subsection{The Survey}
\label{sec:survey}
\subsubsection{Structure}
Our survey is structured in three ordered sections: introduction, items, and subjects' profiling.
In the Introduction section we described the aim of the survey, as well as how data would have been shared and used. 

In the Items section we proposed to subjects tasks related to a cognitive ability of the kind exercised when interpreting a language (e.g., a visual language): knowing the terminology, the syntax, the semantics, and the pragmatics of each DV observed. Our survey consists of eight DVs and four items per DV, thus leading to a total of 32 items.

Items are of the following three \textbf{question types}\footnote{the full survey is available at \url{https://osf.io/ebz8c/overview?view_only=70917ec3404e438ea8529cbe5517b4a8}}:

\begin{enumerate}
\item \textit{Name}: we ask for the name of the DV; 
\item \textit{Function}: we ask questions related to the functional and purpose of the DV, e.g., what the DV represents in terms of data patterns illustrated there or what the DV was usually exploited for, e.g., to show the trend in time, to cluster data, and the like;
\item \textit{Content}: we ask for the extraction of a piece of information from the DV, e.g., the number, the percentage or the label corresponding to the information required in the question posed. 
\end{enumerate}
Each of the eight DV has one \textit{Name}, two \textit{Function} and one \textit{Content} type of question.
The type \textit{Name} is an open question; \textit{Function} and \textit{Content} are multiple-choice, including one correct and three incorrect answers.

Regarding subjects' profiling, to collect data related to the subjects that could act as a proxy of their DVL, we asked about: age, gender, country of birth, education, expertise and experience in data visualization. These subjects' profiling questions were asked after the 32 items to avoid that further fatigue would impact on items responses.

\subsubsection{Randomization}
We define a \textbf{session} as the group of four items related to the same DV. We randomized DVs across subjects; in our terminology, \textit{session x} refers to the group of four items related to the \textit{xth DV} presented to a subject. We did not randomize the order of items within a DV to avoid asking simple questions after hard questions.


\subsubsection{Validation}
The survey items were selected through a rigorous methodology validated in \cite{LocoroGoliaFalessi2025}.

\subsubsection{Population}
Given a budget of € 15,000.00, assuming the survey would take about 1 hour and that a fair payment to respondents would be approximately € 13 per hour, we collected valid responses from 1,083 respondents, yielding 34,656 item responses. We excluded answers from respondents who abandoned the survey, timed out, or responded too quickly (i.e., less than 15 minutes). We excluded non-English speakers and individuals with a PhD from the targeted population. 

\subsection{\rqone}
The dependent variable of this research question is HIC. The independent variables are the question type and the session (see Section \ref{sec:survey}), which act as a proxy for humans' fatigue.

\subsection{\rqtwo}
The dependent variable of this research question is P-HIC accuracy as measured via Area Under the Curve and Kappa.
\begin{itemize}
    \item  \textit{Area Under the Receiver Operating Characteristic Curve} (\textbf{AUC})~\citep{roc} is the area under the curve of true positives rate versus false positive rate, which is defined by setting multiple thresholds. AUC has the advantage of being threshold independent  \cite{DBLP:journals/tse/LessmannBMP08}. We decided to avoid metrics such as Precision,  Recall and F1, since they are threshold dependent.
    
    \item \textit{Cohen’s Kappa}  (\textbf{kappa}) is a statistic that assesses the classifier’s performance against random guessing \cite{Cohen1960Kappa}. It is a chance-corrected measure of agreement for nominal (categorical) labels. In this work, we use Cohen's $\kappa$ as a chance-corrected analogue of accuracy to quantify whether the predictive performance of a classifier is better or worse than what would be expected from a random predictor with the same class prevalence. Specifically, $\kappa$ measures the extent to which the observed agreement between predicted and true class labels exceeds the agreement expected by chance under independence while preserving the marginal label distributions. Consequently, $\kappa=1$ denotes perfect prediction, $\kappa=0$ indicates performance indistinguishable from random guessing at the level implied by the class marginals, and $\kappa<0$ signals systematic disagreement, i.e., performance worse than the corresponding chance baseline. Compared with raw accuracy, $\kappa$ therefore provides a more informative assessment in the presence of class imbalance, because it explicitly subtracts the agreement attributable to prevalence alone \cite{Cohen1960Kappa}.
\end{itemize}

In this study, our positives are the incorrect subjects' answers, and the negatives are the correct subjects' answers; i.e., a \textit{true-positive} is when P-HIC predicts an incorrect answer and the subject provided an incorrect answer.

We now elaborate on our P-HIC approach, which will be used to measure these dependent variables.


\subsubsection{Features}
\label{sec:feature}
The first step to make a prediction is to define and then measure the inputs of the prediction model, aka, \textbf{features}. In this context, we sought information that could be correlated with HIC and is measurable before the subject interacts with the item, i.e., before sees the DV. This work takes into consideration 22 features, grouped in three families: 

\begin{itemize}
    \item \textit{Item difficulty}: this family consists of two features. The first feature, \textit{ExpertDifficulty}, provides, for each item, the median score based on the judgments of seven raters, recruited from scholars at two universities (the authors) and external advisors with different expertise in DV~\cite{LocoroGoliaFalessi2025}. 
The second feature is \textit{RaschDifficulty}, obtained by applying the Rasch Model (\textbf{RM}). The RM is a measurement model that transforms raw scores into linear and reproducible measures~\cite{Rasch1960}. It specifies the probability that a person gives a correct response to an item as a function of the person's level of a latent trait (person parameter) and the item difficulty (item parameter). A distinctive feature of the RM is the separability of person and item parameters. Separability implies that person parameters are conditioned out during item calibration, yielding sample-independent item estimates, and item parameters are conditioned out during person measurement, yielding test-independent person estimates. The RM assumes unidimensionality and local independence; when the data fit the model, the resulting measures are objective, and their unit of measurement is the logit, which has interval-scale properties~\cite{WrightMasters1982, BondFox2015}. In the present analysis, the parameters involved in the RM were estimated using the joint maximum likelihood estimation method~\cite{WrightMasters1982} under the constraint that the sum of the item difficulty parameters was equal to 0.0 logits. 
    The estimated difficulty of the items, obtained using the software Winsteps 5.4~\cite{WINSTEP}, ranged between -2.38 and 2.36 logits with zero mean and 1.09 standard deviation. To avoid data leakage, the RM was computed for each item-subject combination by considering the answers provided to the item by all subjects except the current one.    
    
    \item \textit{Human Profile}: this family consists of 18 features and relates to the characteristics of the subject, providing an answer to the item, as provided by the subject itself and possibly related to the DVL of the subject. These features include age, gender, country of birth, expertise and experience in data visualization and education level as asked in the human profile section of the survey.

    \item \textit{Human Performance}: this family consists of two features, \textit{PercCorrect} and \textit{MedianDifficulty}, which depend on the current data. 
    The survey responses were arranged into 32 datasets, following the sequence of the items. For example, dataset 10 contains the responses given by the subjects to the item administered as their 10th item. 
    For each dataset, starting from the second one, we computed \textit{PercCorrect} as the percentage of correct responses provided by the subject to previous items. This indicator can be seen as an approximation of the subject's DVL level. Accordingly, for dataset 10, \textit{PercCorrect} is the percentage of correct responses provided to the previous nine items.     
    The second feature is \textit{MedianDifficulty}, which was obtained as the median \textit{RaschDifficulty} of the items, on which the \textit{PercCorrect} is calculated. This feature serves as a counterweight to \textit{PercCorrect}. In fact, if \textit{PercCorrect} assumes a high value, it matters whether this value is derived from a set of items that are all easy, all difficult, or a mix of both.
   
\end{itemize}

\subsubsection{Models}
\label{Subsec.Models}
P-HIC consists of three well-established ML models: 

\begin{itemize}
    \item \textit{Logistic Regression}  (\textbf{LR}): it is a parametric statistical model used to estimate the probability of an event of interest, in which the associated log-odds are modeled as a linear combination of multiple independent predictors \cite{HosmerLemeshowSturdivant2013}. Model parameters are estimated by maximizing the log-likelihood function, typically via a Newton–Raphson algorithm. Categorical independent predictors with k categories are replaced by k-1 dummy variables, with one category serving as the reference.
 
    \item \textit{Multi Layer Perceptron} (\textbf{MLP}): it implements a feed-forward multi-layer perceptron trained with backpropagation for supervised classification (and regression when the class is numeric). Network topology can be specified explicitly or generated automatically via a simple heuristic for the hidden layers. Training behavior can be controlled through standard gradient-based hyperparameters, such as the learning rate and momentum, with optional learning-rate decay to improve stability. All hidden units use sigmoid activations; output units are sigmoid for nominal classes, whereas for numeric targets the outputs are linear (unthresholded) and the target may be optionally normalized internally. The implementation supports practical training controls, including random seeding (for weight initialization and data shuffling), early stopping based on validation set deterioration (with configurable validation split and patience threshold), or fixed epoch budgets, and an optional resume mode to continue training beyond the initial run. Preprocessing options include nominal-to-binary conversion and attribute normalization (including scaling binary indicators to a symmetric range), which may improve convergence \cite{DBLP:books/lib/Bishop07}.

    \item \textit{Random Forest} (\textbf{RF}): it constructs an ensemble of randomized decision trees following Breiman’s RF methodology. Each tree is trained on a bootstrap sample (bag) drawn from the training set, and at each split, only a random subset of attributes is considered, thereby reducing correlation between trees and improving generalization. Predictions are aggregated across trees (e.g., by majority vote for classification). The implementation provides controls for reproducibility (random seed), ensemble size (number of trees), bootstrap sampling rate (bag size percentage), and tree complexity (maximum depth, with an option for unlimited depth). It can compute out-of-bag (OOB) error estimates—optionally storing OOB predictions and emitting additional OOB statistics—and can compute attribute importance via mean impurity decrease \cite{DBLP:journals/ml/Breiman01}.
\end{itemize}
We used these models in isolation and with the default parameters in WEKA 3.9.6\footnote{https://sourceforge.net/projects/weka/files/weka-3-9/3.9.6/}.

\subsubsection{Feature Selection}
A prediction model can be used with and without feature selection. As feature selection we used two complementary filter-based selectors: (i) correlation-based feature selection (\textbf{CFS}) with CfsSubsetEval and GreedyStepwise, which targets parsimonious subsets by explicitly trading off high predictor–class association against low inter-predictor redundancy, as originally formalized and empirically validated in Hall’s CFS framework \cite{DBLP:conf/icml/Hall00} and consistent with the broader relevance–redundancy perspective on feature selection \cite{DBLP:journals/ai/BlumL97,DBLP:journals/jmlr/GuyonE03}; and (ii) univariate ranking with GainRatioAttributeEval, leveraging Quinlan’s gain ratio to obtain an interpretable screening criterion while mitigating the known multi-valued attribute preference of information gain in decision-tree induction \cite{DBLP:books/mk/Quinlan93}.
Feature selection was integrated into the learning pipeline using WEKA’s AttributeSelectedClassifier, which recomputes the selected feature set using only the training partition of each cross-validation fold and then applies the identical projection to the corresponding test partition before fitting the final learner, thereby preventing information leakage during model assessment.  Embedding attribute selection inside the resampling loop is methodologically important because performing feature selection (or any model selection step) outside cross-validation can lead to optimistically biased error estimates; this risk and the need for strictly fold-wise (or nested) selection are well documented in the model selection and performance evaluation literature \cite{DBLP:conf/ijcai/Kohavi95,DBLP:journals/bmcbi/VarmaS06,DBLP:journals/jmlr/CawleyT10}. The resulting pipeline follows established best practices in applied machine learning experimentation and is directly supported by WEKA’s design for reproducible end-to-end evaluation \cite{DBLP:books/sp/datamining2005/FrankHHKP05,DBLP:books/lib/WittenFH11}.

\subsubsection{Measurement Procedure}
\label{sec:MeasurementProcedure}
To measure accuracy, i.e., our dependent variable, we opted for a ten-time ten-fold validation as suggested by Witten et al. \cite{DBLP:books/sp/datamining2005/FrankHHKP05,DBLP:books/lib/WittenFH11}. This was done for each classifier, with and without feature selection, in isolation.
We created 32 datasets, each related to the subjects’ responses provided for the $i-th$ item ($i = 1, \dots, 32$). The ten-time ten-fold validation was applied to each dataset in isolation; thus each dataset was predicted 100 times.
We note that, given the design of the survey, the first dataset was related to the question \textit{Name}, the 1,083 rows are related to the 1,083 different subjects answering the Name of a DV (among the eight DV randomized across subjects).

\subsection{\rqthree}
\label{sec:RQ3Design}
In order to determine which of the 22 features (independent variables) described in Section~\ref{sec:feature} are relevant for P-HIC, we used two metrics (dependent variables): Information Gain and accuracy achieved by a specific set of features.

Information Gain is as a ranking criterion that quantifies the predictive utility of each feature with respect to the class by using Quinlan’s Gain Ratio, i.e., an information-gain score normalized by the attribute’s intrinsic information. This normalization is specifically intended to mitigate the well-known tendency of plain information gain to favor features with many distinct values, thereby yielding a more balanced and comparable univariate relevance score across heterogeneous predictors \cite{DBLP:books/mk/Quinlan93}. Thus, Gain Ratio quantifies how much knowing a feature reduces uncertainty about the class, after normalizing by the feature’s intrinsic information; larger values therefore indicate stronger univariate association with the class and, in a ranking setting, greater expected utility for prediction. Values close to zero imply that the feature provides little or no class-discriminative information, either because the class distribution is similar across the feature’s values, or because any reduction in class entropy is negligible after normalization. There is no universal absolute threshold that separates “good” from “bad” Gain Ratio scores: interpretation is primarily comparative within a given dataset, where attributes are ordered by score and selection is typically based on the top-ranked variables or on performance-driven cutoffs downstream. \cite{DBLP:books/mk/Quinlan93}. The Gain Ratio is computed on the training set on each of the 100 runs (ten-time ten-folds) on each of the 32 datasets created for RQ2.

Following a practice that has been adopted in related empirical contexts \cite{DBLP:journals/ese/FalessiLCEC23,DBLP:journals/tse/FalessiRGC20}, we analyze predictive accuracy not only when using the full feature set, but also when restricting the classifier to individual, semantically coherent groups of features. This ablation-style comparison estimates the marginal contribution of each group to the predictive signal. If a model trained on a single group achieves accuracy comparable to that obtained with all groups combined, then the remaining groups provide limited additional information beyond what is already captured by that group, suggesting redundancy or weak incremental predictive value. Conversely, substantial performance drops when excluding a group indicate that the omitted features encode complementary information that is relevant to prediction. For computing the accuracy, we used the same setup as for RQ2, i.e., a ten-time ten-fold cross-validation for each of the 32 individual datasets, and as a classifier, we used the best classifier identified for RQ2. 

We grouped the features already described in Section\ref{sec:feature} under the following sub-groups:
\begin{itemize}
    \item \textit{Human Profile}: it includes the 18 features related the characteristics of the human such as \textit{age}. This set of features is interesting to analyze in this context since it represents the accuracy that is achievable by P-HIC without considering any information about the question under prediction. 
    \item \textit{Human Profile and Performance}: it includes the "Human Profile" and "Human Performance" (e.g., \textit{PercCorrect}) feature groups. This is interesting to analyze in this context since it represents the accuracy that is achievable by P-HIC without considering any measure of items' difficulty. 
    \item \textit{Only Rasch}: it includes the Rasch score associated with the item, i.e., \textit{RaschDifficulty}. This is interesting to analyze in this context, since it represents the accuracy achievable by P-HIC without considering any information about the subject (profile or performance) being predicted. 
    \item \textit{All}: it includes all the features described in Section\ref{sec:feature}. This is interesting to analyze in this context since it represents the P-HIC accuracy upper bound.  
\end{itemize}

\section{Study Results Analysis}
\label{sec:results}

\subsection{\rqone}
\label{sec:rq1res}
\begin{figure}[t]
  \centering
   \includegraphics[width=1\columnwidth]{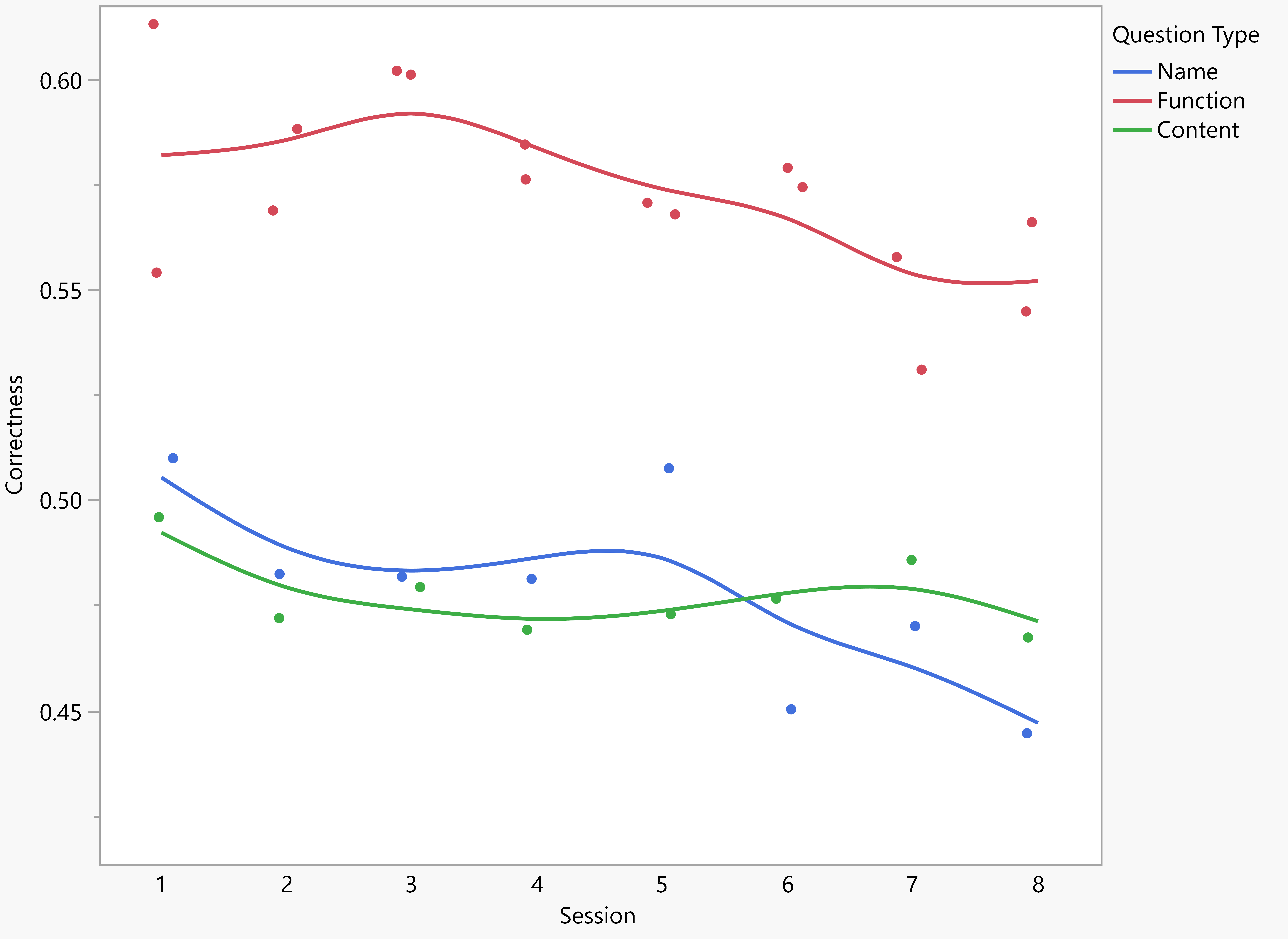}
   \caption{Average HIC on different question types (colors) and sessions (horizontal axis). } 
   \label{Figure:AccuracyHumans}
\end{figure}

Figure~\ref{Figure:AccuracyHumans} reports the average HIC across subjects, related to question type (color) and sessions (horizontal axis). This result shows that the \textit{Function} type resulted easier to answer than the \textit{Content} and the \textit{Name} types. We can also observe different dynamics in the average HIC of the three types of questions across the eight sessions. The average HIC of \textit{Name} question appears to decrease over time, and this is confirmed by applying McNemar’s chi-squared test for paired nominal data to the first and last sessions. The test highlights a significant difference in the distribution of average HIC between sessions one and eight (p-value < .001). In contrast, the average HIC of the \textit{Content} question remains almost constant.
Regarding the \textit{Function} question, a decline in the average HIC is observable in the figure; however, this trend cannot be statistically validated due to the random nature of the study design for this type of question.

\subsection{\rqtwo}
\label{sec:rq2sec}
Table \ref{table:RQ1ClassifiersComparison} reports three statistics used to determine which of the three models introduced in Section~\ref{Subsec.Models} outperforms the others. The first statistic, $\% Best$, indicates the percentage of times a model was the best-performing one across the 32 datasets, whereas $Median(AUC)$ and $Median(kappa)$ report the median values computed over the 32 datasets and the 100 runs per dataset. Each of the three models is evaluated with and without a prior feature selection step. These results show that the most accurate model is the LR model in combination with feature selection. Due to this overperformance, in the remainder of the analysis, solely this model with feature selection is considered. To support replication and alternative analyses, we report the results of all classifiers, with and without feature selection, of each of 100 runs on each of the 32 datasets\footnote{the full results are available at \url{https://osf.io/ebz8c/overview?view_only=70917ec3404e438ea8529cbe5517b4a8}}.

\begin{table}[]
\caption{Models accuracy with and without feature selection.}
	\label{table:RQ1ClassifiersComparison}
\begin{tabular}{|c|c|c|c|c|}
\hline
\textbf{Classifier} & \textbf{FS} & \textbf{\% Best} & \textbf{Median(AUC)} & \textbf{Median(kappa)} \\ \hline
LR            & no          & 0\%              & 0.686                & 0.287                  \\ \hline
LR            & yes         & 81\%             & 0.724                & 0.319                  \\ \hline
MLP                 & no          & 0\%              & 0.660                & 0.179                  \\ \hline
MLP                 & yes         & 13\%             & 0.718                & 0.294                  \\ \hline
RF        & no          & 3\%              & 0.684                & 0.269                  \\ \hline
RF        & yes         & 3\%              & 0.681                & 0.260                  \\ \hline
\end{tabular}
\end{table}

\begin{figure}[t]
   \centering
   \includegraphics[width=1\columnwidth]{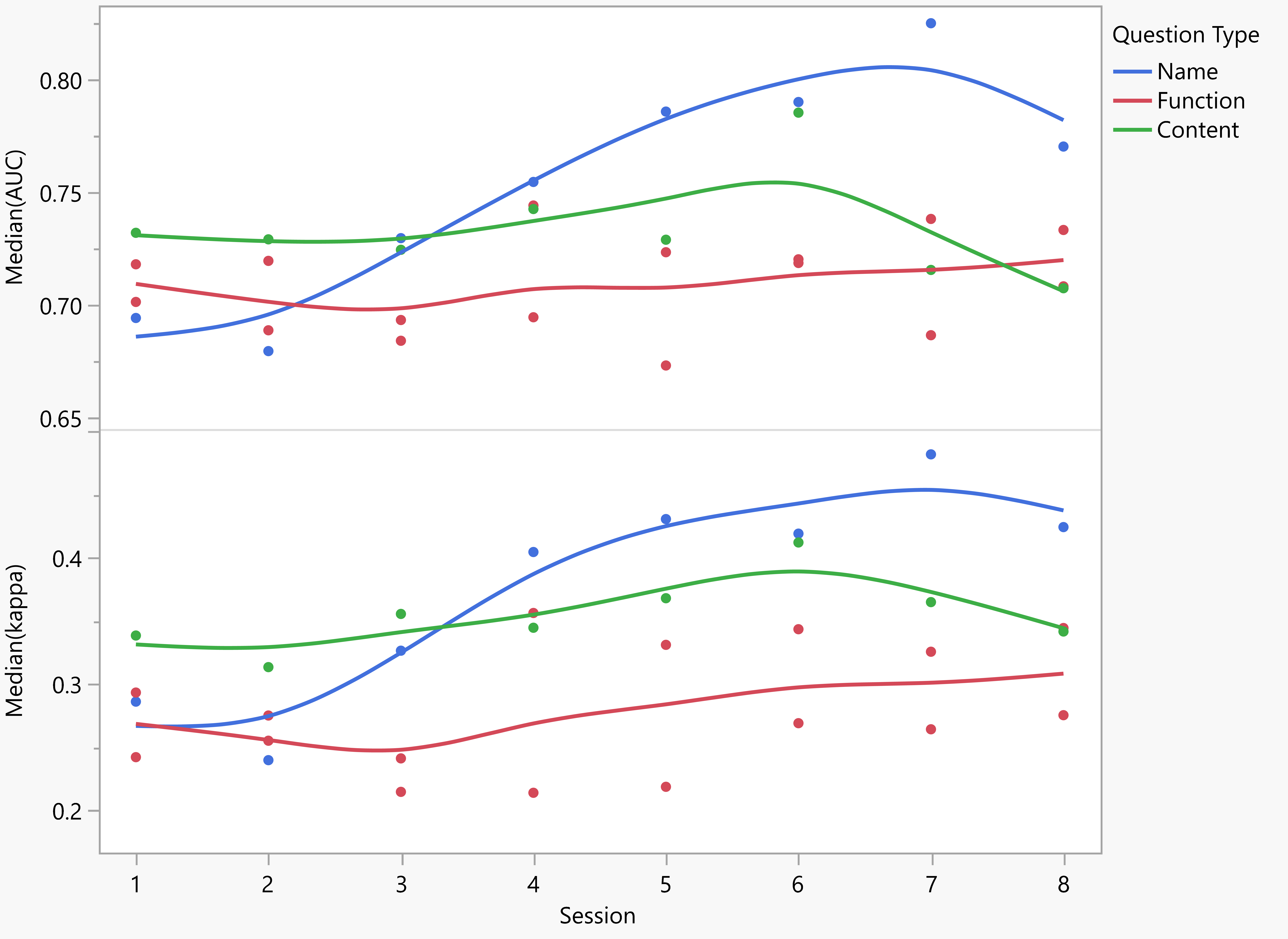}
   \caption{Accuracy (vertical axis) of P-HIC (LR model) along different sessions (horizontal axis), by question type (color) using all features. } 
   \label{Figure:AccuracyBySessionAndQuestionType}
\end{figure}

Figure~\ref{Figure:AccuracyBySessionAndQuestionType} reports the median accuracy, in terms of AUC and kappa, in the 100 predictions for each session, of P-HIC using LR model and all features. The colors represent the question type, the horizontal axis indicates the session. 
This result shows that the median AUC across the 100 runs is pretty stable across sessions for the \textit{Function} and \textit{Content} type of question.
Instead, for the \textit{Name} type of question the median AUC increases until the seventh session, with an $18\%$ increase from session one to session seven.

\subsection{\rqthree}

\begin{figure*}[t]
    \centering
    \includegraphics[width=1\textwidth]{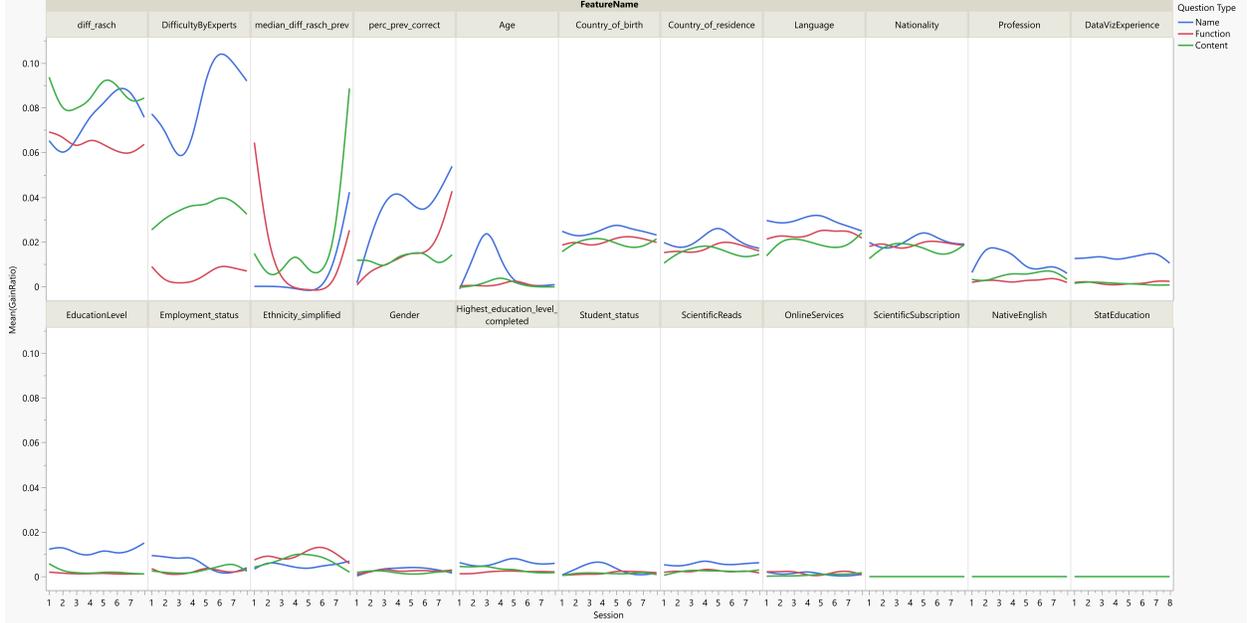}
   \caption{Importance (vertical axis) of each prediction feature (columns) in different sessions (horizontal axis) for different types of questions (color).} 
    \label{Figure:InfoGainScatter}
\end{figure*}

Figure~\ref{Figure:InfoGainScatter} reports the importance in terms of mean Gain Ratio for each feature (column) in specific sessions (horizontal axis), by question type (color), over the 100 runs. The figure shows each feature in order of importance, from the top left to the bottom right. This result shows that the \textit{RaschDifficulty} is by far the most important feature for P-HIC. This feature is followed by the difficulty score given by experts, which seems to support in particular the P-HIC related to the question type \textit{Name} over the other two question types. The third relevant feature is
\textit{PercCorrect}, i.e. the percentage of previous correct answers. Its importance increases with the session number, especially for the \textit{Name} question type, but also for the \textit{Function} question type. Moreover, the median difficulty computed over the previous responses based on Rasch score, i.e., \textit{MedianDifficulty} is worth mentioning. For this feature, we observe higher performance in the first half of the sessions for the question type \textit{Function} and higher performance in the second half of the sessions for the question type \textit{Content}. Other relevant features are \textit{CountryOfBirth}, \textit{CountryOfResidence}, \textit{Language}, and \textit{Nationality}, which seem to be equally important for the three question types, and outperform features like \textit{Age}, \textit{Profession}, and \textit{DataVizExperience}.

\begin{figure*}[t]
    \centering
    \includegraphics[width=1\textwidth]{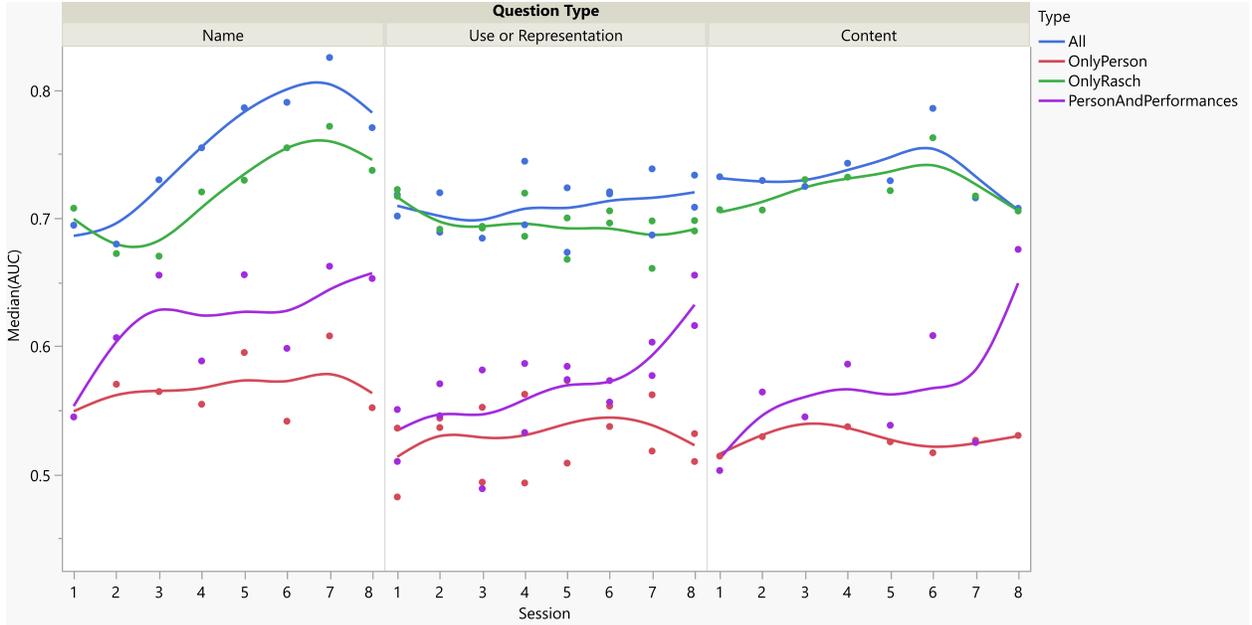}
   \caption{Accuracy (vertical axis) of P-HIC (LR model) along different sessions (horizontal axis) using specific feature groups (color) by question type (column).} 
    \label{Figure:AccuracyByFeaturesGroups}
\end{figure*}

Figure~\ref{Figure:AccuracyByFeaturesGroups} reports the accuracy of P-HIC (best model) along different sessions (horizontal axis) using specific feature groups (color) as defined in Section~\ref{sec:RQ3Design} by question type (column). This result shows that the accuracy of P-HIC using \textit{Only Rasch} is very similar to \textit{All}; this means that most, if not all, prediction information is provided by the difficulty of the item under prediction, i.e., \textit{RaschDifficulty}, rather than the characteristics of the subject answering the item. According to Figure~\ref{Figure:AccuracyByFeaturesGroups} the type of question does not impact the importance of the feature group; i.e., \textit{RaschDifficulty}, is by far the most important feature group followed by \textit{Human Profile and Performance} and \textit{Human Profile}.

\section{Discussion}
\label{sec:discussion}

This study investigates the interpretation correctness through a survey with DV items, and the prediction of interpretation correctness, before exposure, i.e., P-HIC, using a combination of user traits, item difficulty indicators, and contingent performance features, and by leveraging standard machine learning models. Our results support the feasibility of this goal and provide insights into both the behavioral dynamics of visualization interpretation and the predictive factors that drive correctness.

\subsection{\rqone}
A key central implication of our findings concerns the temporal dynamics of HIC across sessions, which provide evidence of fatigue-like effects and, importantly, offer practical guidance on how many items can be reasonably administered in a single assessment. The analysis of raw HIC across sessions shows that HIC trends differ by question type: \textit{Function} items (Use/Representation) were easier overall, while \textit{Name} and \textit{Content} items were harder, with \textit{Name} exhibiting a statistically significant decline across sessions (McNemar test between sessions 1 and 8, $p < .001$). By contrast, \textit{Content} HIC remained substantially stable.  This pattern suggests that some interpretation tasks may be more sensitive to cognitive depletion or attentional decay than others. In particular, \textit{Name} questions require retrieving technical terminology (i.e., a lexicon of visualization types), which may be affected by fatigue or reduced motivation, whereas \textit{Content} extraction tasks appear to rely on more stable perceptual decoding processes. This finding has direct consequences for assessment design: the number of items and their ordering should be considered carefully, and test designers may use such trends to identify “drop-off points” beyond which additional items yield diminishing diagnostic value.

The stability of the predictive accuracy over sessions (see RQ2) suggests that the proposed modeling approach is robust to these evolving user dynamics. In other words, although HIC changes over time, the predictive model remains relatively stable across sessions and question types (see Figure~\ref{Figure:AccuracyBySessionAndQuestionType}); this implies that the chosen feature set captures a meaningful portion of the relevant behavioral signal, including evolving performance.

Overall, the above findings indicate that fatigue bias is an informative phenomenon rather than merely a confounding factor to control for; fatigue bias identification helps identify how many items are appropriate to administer and demonstrates that predictive models can incorporate this factor, given that contingent performance indicators, e.g., \textit{PercCorrect}, increase in importance as sessions progress (see RQ3).

\subsection{\rqtwo}
Regarding prediction performance, we found that the best-performing model was the LR model combined with feature selection, which is the best model for 26 of the 32 dataset ($\%Best = 81$), achieving a median AUC of 0.724 and median $\kappa$ of 0.319, outperforming both RF and MLP variants under the same evaluation protocol (10$\times$10-fold validation over 32 datasets). This result is notable because it suggests that the P-HIC task benefits from interpretable linear decision boundaries, likely because many predictive features (e.g., Rasch difficulty, expert difficulty, cumulative performance indicators) encode relatively monotonic relationships with HIC. In contrast, more complex models, such as MLP, did not yield consistent superiority and performed best only when combined with feature selection. The strong performance of Logistic Regression model also has practical implications: it suggests P-HIC can be integrated into assessment pipelines with low computational costs, good transparency, and easier interpretability for feature-driven customization.

Moreover, predictive performance remains relatively stable across sessions for the \textit{Function} and \textit{Content} type of question (see Figure~\ref{Figure:AccuracyBySessionAndQuestionType}), which is encouraging for real-world deployment scenarios where users’ engagement and performance may fluctuate \cite{beschi2025ai}. Instead, the predictive performance of the \textit{Name} type of question increases as the number of items answered by the subjects grows. One possible explanation is the increasing precision of the feature \textit{PercCorrect}, which serves as a proxy for the subject’s DVL level; this feature appears to play a more important role for this type of question, which represents a more challenging task, as shown in \cite{LocoroGoliaFalessi2025}.

\subsection{\rqthree}
The analysis of feature importance (see Figure~\ref{Figure:InfoGainScatter}) and accuracy of group of features (see Figure~\ref{Figure:AccuracyByFeaturesGroups}) provides a clear answer to where the predictive signal comes from, the question of what drives predictive accuracy in P-HIC. Most importantly, \textit{RaschDifficulty} is by far the most informative feature across sessions and question types (Figure~\ref{Figure:InfoGainScatter}), confirming that objective psychometric modeling is strongly aligned with interpretable HIC.

According to Figure~\ref{Figure:InfoGainScatter}, the second most informative feature is the difficulty rating assigned by experts, and its predictive contribution is especially pronounced for \textit{Name} items, suggesting that experts are particularly good at anticipating the difficulty of terminology/recognition tasks.

Beyond static difficulty, performance history features provide additional predictive power and are particularly important for modeling dynamic changes across sessions. Specifically:

\begin{itemize}
\item \textit{PercCorrect} (percentage of correct responses given before the current item) becomes increasingly useful over time, especially for \textit{Name} and \textit{Function} items, suggesting that individual contingent performance becomes more informative as the test progresses.
\item \textit{MedianDifficulty} of previously answered items emerges as a strong feature for \textit{Function} items in early sessions and for \textit{Content} items in later sessions.
\end{itemize}

These results are important because they support our underlying “beyond literacy” hypothesis: HIC is not only a function of stable skill and item difficulty, but also of interaction-dependent factors, such as momentum, fatigue, overconfidence, or adaptation. The progressive increase in relevance of performance-related features suggests that P-HIC models can continuously recalibrate predictions as they observe user responses, which is a key requirement for adaptive assessment and personalized visualization delivery.

Interestingly, demographic and contextual variables (e.g., \textit{CountryOfBirth}, \textit{CountryOfResidence}, \textit{Language}, \textit{Nationality}) also rank higher than more commonly assumed predictors such as \textit{Age}, \textit{Profession}, and \textit{DataVizExperience}. This result can be interpreted in multiple ways: it may indicate that cross-cultural familiarity with specific visualization conventions influences HIC; alternatively, it may reflect latent factors such as educational system differences, exposure to data-driven communication, or language-related interpretation biases in items. While these findings are exploratory, they underscore the need for future work to examine fairness and cultural transferability of visualization literacy assessments.

Finally, the ablation analysis (see Figure~\ref{Figure:AccuracyByFeaturesGroups}) provides a practical guideline for deployment: although using all features yields the best overall accuracy, the \textit{Only Rasch} feature group alone is already highly predictive, confirming that psychometric difficulty is a foundational driver. 

\subsection{Advancements from  Related Work}

Research on visualization literacy and visualization comprehension  \cite{locoro2021visual,beschi2025ai,verma2024evaluating,Lee2017551} has introduced standardized visualization literacy tests, task taxonomies, and empirical analyses of how users interpret charts and graphs, emphasizing constructs such as graphical fluency, visual encoding effectiveness, and task-dependent comprehension. This literature has been highly influential in identifying what makes visualizations difficult to interpret, but it has predominantly focused on descriptive or explanatory analyses of correctness after user responses are collected.

The present work proposes a predictive perspective that complements and extends those prior contributions. By formulating visualization interpretation as an item-level binary prediction task, this study moves beyond static literacy assessment and demonstrates that correctness can be anticipated before exposure to an item. This represents a methodological shift that aligns visualization research with predictive modeling practices, while remaining grounded in visualization-specific theory and tasks.

Prior studies such as \citet{verma2025measuring,chang2025tell} have emphasized the role of stable user characteristics, including visualization experience, education level, and domain expertise, as determinants of comprehension. Our findings partially support this view but also challenge its centrality. Across question types and sessions, Rasch-based item difficulty emerges as the most informative predictor of correctness, consistently outperforming self-reported experience and demographic variables. This result bridges visualization research with psychometric modeling, and demonstrates their applicability and value within DV research.

At the same time, adaptive assessment and personalization, often discussed as future opportunities rather than fully operationalized systems \cite{cui2023adaptive}. Our results provide concrete empirical support for these directions. The increasing importance of performance-history features across sessions mirrors findings from \citet{halkiopoulosGkintoni2024}, where adaptive testing and learner modeling rely on continuously updated estimates of user state. By showing that similar mechanisms improve prediction of visualization interpretation correctness, this work extends adaptive assessment concepts to visualization literacy, an area where they have so far seen limited empirical validation.

Although visualization task taxonomies were introduced in \cite{Ge2024Visualization}, the distinction between naming, content extraction, and functional reasoning tasks, was rarely examined in their role of affecting predictive modeling. Our feature-importance analysis reveal task-specific predictive patterns, with expert-rated difficulty being particularly relevant for naming tasks and contingent performance features playing a stronger role for other task types. This provides new empirical grounding for task-based theories of visualization comprehension that have been proposed in the literature, but not previously linked to predictive performance.

Finally, from a methodological perspective, recent discussions \cite{lily2025autoethnography} have raised concerns about model transparency, interpretability, and practical deployability. The strong performance of LR model over more complex models in our study contributes to this discussion by showing that theory-informed features paired with interpretable models can yield competitive or superior results. This supports a growing emphasis within these studies on responsible and explainable modeling of human behavior.

\subsection{Implications for Design}
A major outcome of this study is its potential to support adaptive and personalized assessment. Once the most informative features are identified (notably Rasch-based indicators, expert difficulty, and dynamic performance signals), users can be profiled prior to assessment, and items can be selected that maximize measurement efficiency while reducing user burden.

In practical terms, P-HIC enables an adaptive mechanism such as:
\begin{enumerate}
\item estimating the probability that a user will answer an item correctly,
\item avoiding overly difficult items that would lead to misunderstanding or discouragement,
\item prioritizing items that best match the user’s level and maximize diagnostic information,
\item customizing visualization training by selecting items that are challenging but feasible.
\end{enumerate}

This approach aligns with the broader goals of personalized human-data interaction: choosing the right visualization for the right audience at the right time, thereby improving comprehension, reducing cognitive burden, and enabling more effective decision-making.

\subsection{Threats to Validity}
We note that the objective of RQ2 is not to maximize predictive performance with the most recent model families, but is to assess whether P-HIC can predict HIC before interaction, i.e., using only information that is measurable before the subject interacts with the item and to determine which of the proposed prediction features are most informative of HIC (RQ3). For this reason, we deliberately adopt three well-established models as implemented in WEKA (LR, RF, and MLP) as methodological baselines. This conservative choice limits confounding due to extensive model/architecture engineering and it allows observed performance to be attributed primarily to the proposed feature set (e.g., Item difficulty, Person Profile, and Person Performance) rather than to highly tuned model capacity. Moreover, these standard models support transparent analyses of feature importance, which is the goal of RQ3. Future work can replace these robust baselines with more recent architectures once feasibility and the key predictors have been established.

While the results are promising, several limitations should be acknowledged. First, the dataset includes only English-speaking respondents with at most a graduate degree, which limits generalizability to broader populations. Second, prediction relies on item-level difficulty parameters derived from the observed sample, although Rasch was computed carefully in a leakage-free way. Third, the study focuses on HIC and does not yet model response time, uncertainty, or explanation quality, which may be critical in some settings \cite{kempt2024explainable, beschi2025ai}. 

Future work may extend P-HIC by incorporating richer features (e.g., eye-tracking, interaction traces) or by exploring alternative personalization policies (e.g., reinforcement learning, optimal stopping). More importantly, the evidence that language and nationality-related variables carry predictive signal motivates deeper work on cross-cultural validity and equity of visualization literacy testing.

\section{Conclusion}
\label{sec:conclusions}
This paper investigated HIC in a DVL assessment based on DV items and demonstrated the feasibility of P-HIC, i.e., predicting correctness before exposing the human to the target item. Across 1,083 participants and 34,656 item responses, HIC varies systematically across question types and sessions: Function items are easier overall, whereas Name and Content items are harder, with Name exhibiting a statistically significant decline between early and late sessions, consistent with fatigue-like effects. Building on these behavioural results, we evaluated three baseline machine-learning models for P-HIC using a ten-time ten-fold validation protocol across 32 datasets. LR combined with feature selection yielded the strongest and most consistent predictive performance (median AUC 0.72; median kappa 0.32), indicating that P-HIC can be achieved with models that remain transparent and computationally lightweight.

From a research perspective, our results provide empirical support for the “beyond literacy” hypothesis: HIC is not solely explained by stable skill and item difficulty, but also by interaction-dependent dynamics that evolve over the course of an assessment. Feature analyses show that \textit{RaschDifficulty} is the dominant predictor across sessions and question types, followed by \textit{ExpertDifficulty} and contingent performance indicators such as \textit{PercCorrect}, whose relevance increases over time. This highlights a methodological pathway for integrating psychometric modeling with predictive approaches to study temporal effects (e.g., fatigue bias) and to inform the design of adaptive DVL instruments that continuously recalibrate predictions as additional responses are observed.

From a practical perspective, P-HIC directly motivates runtime selection of DV items tailored to an audience. The observed combination of robust accuracy and interpretability supports deployment in assessment and training pipelines where transparency and low computational cost are important. Concretely, P-HIC can be used to (i) estimate the probability of correct interpretation, (ii) avoid overly difficult items that may lead to misunderstanding or discouragement, (iii) prioritize items that maximize diagnostic information, and (iv) personalize training by selecting items that are challenging but feasible, thereby reducing cognitive burden and improving decision-making support.

Future work should validate these findings on additional DV item sets and broader populations, and extend P-HIC by incorporating richer signals (e.g., response time, uncertainty, explanation quality, eye-tracking, interaction traces) and by evaluating alternative personalization policies (e.g., reinforcement learning, optimal stopping) for end-to-end adaptive assessment. Finally, the predictive contribution of language- and nationality-related variables calls for focused investigation of cross-cultural validity and equity of DVL testing.

\section*{Acknowledgments}
``Research funded by the European Union – Next-Generation EU - PRIN 2022 D.D. 104 of 02-02-2022, Mission 4 Component 1
CUP master D53D23008690006, CUP E53D23008080006, project name: ``Characterizing and Measuring Visual Information Literacy'' ID 2022JJ3PA5.''






\begin{thebibliography}{38}
\providecommand{\natexlab}[1]{#1}
\providecommand{\url}[1]{#1}
\csname url@samestyle\endcsname
\providecommand{\newblock}{\relax}
\providecommand{\bibinfo}[2]{#2}
\providecommand{\BIBentrySTDinterwordspacing}{\spaceskip=0pt\relax}
\providecommand{\BIBentryALTinterwordstretchfactor}{4}
\providecommand{\BIBentryALTinterwordspacing}{\spaceskip=\fontdimen2\font plus
\BIBentryALTinterwordstretchfactor\fontdimen3\font minus \fontdimen4\font\relax}
\providecommand{\BIBforeignlanguage}[2]{{%
\expandafter\ifx\csname l@#1\endcsname\relax
\typeout{** WARNING: IEEEtranN.bst: No hyphenation pattern has been}%
\typeout{** loaded for the language `#1'. Using the pattern for}%
\typeout{** the default language instead.}%
\else
\language=\csname l@#1\endcsname
\fi
#2}}
\providecommand{\BIBdecl}{\relax}
\BIBdecl

\bibitem[Locoro et~al.(2021)Locoro, Fisher, and Mari]{locoro2021visual}
A.~Locoro, W.~P. Fisher, and L.~Mari, ``Visual information literacy: Definition, construct modeling and assessment,'' \emph{IEEE access}, vol.~9, pp. 71\,053--71\,071, 2021.

\bibitem[Beschi et~al.(2025{\natexlab{a}})Beschi, Falessi, Golia, and Locoro]{beschi2025characterizing}
S.~Beschi, D.~Falessi, S.~Golia, and A.~Locoro, ``Characterizing data visualization literacy for standardization: A systematic literature review,'' \emph{IEEE Access}, 2025.

\bibitem[Locoro and Lavazza(2025)]{locoro2025modeling}
A.~Locoro and L.~Lavazza, ``Modeling interaction patterns in visualizations with eye-tracking: A characterization of reading and information styles,'' \emph{Future Internet}, vol.~17, no.~11, p. 504, 2025.

\bibitem[Beschi et~al.(2025{\natexlab{b}})Beschi, Fogli, Gargioni, and Locoro]{beschi2025ai}
S.~Beschi, D.~Fogli, L.~Gargioni, and A.~Locoro, ``Ai-supported eud for data visualization: An exploratory case study,'' \emph{Future Internet}, vol.~17, no.~8, p. 349, 2025.

\bibitem[Lee et~al.(2017)Lee, Kim, and Kwon]{Lee2017551}
S.~Lee, S.-H. Kim, and B.~C. Kwon, ``Vlat: Development of a visualization literacy assessment test,'' \emph{IEEE Transactions on Visualization and Computer Graphics}, vol.~23, no.~1, p. 551 – 560, 2017.

\bibitem[Pandey and Ottley(2023)]{Pandey20231}
S.~Pandey and A.~Ottley, ``Mini-vlat: A short and effective measure of visualization literacy,'' \emph{Computer Graphics Forum}, vol.~42, no.~3, p. 1 – 11, 2023.

\bibitem[Page et~al.(2021)Page, McKenzie, Bossuyt, Boutron, Hoffmann, Mulrow, Shamseer, Tetzlaff, Akl, Brennan, et~al.]{page2021prisma}
M.~J. Page, J.~E. McKenzie, P.~M. Bossuyt, I.~Boutron, T.~C. Hoffmann, C.~D. Mulrow, L.~Shamseer, J.~M. Tetzlaff, E.~A. Akl, S.~E. Brennan \emph{et~al.}, ``The prisma 2020 statement: an updated guideline for reporting systematic reviews,'' \emph{Bmj}, vol. 372, 2021.

\bibitem[Cui et~al.(2024{\natexlab{a}})Cui, Ge, Ding, Yang, Harrison, and Kay]{cui2023adaptive}
Y.~Cui, L.~W. Ge, Y.~Ding, F.~Yang, L.~Harrison, and M.~Kay, ``Adaptive assessment of visualization literacy,'' \emph{IEEE Transactions on Visualization and Computer Graphics}, vol.~30, no.~1, pp. 628--637, 2024.

\bibitem[Cui et~al.(2024{\natexlab{b}})Cui, Ge, Ding, Harrison, Yang, and Kay]{Cui-Promises-Pitfalls}
Y.~Cui, L.~W. Ge, Y.~Ding, L.~Harrison, F.~Yang, and M.~Kay, ``Promises and pitfalls: Using large language models to generate visualization items,'' \emph{IEEE Transactions on Visualization and Computer Graphics}, pp. 1--11, 2024.

\bibitem[Verma and Fan(2025)]{verma2025measuring}
A.~Verma and J.~E. Fan, ``Measuring and predicting variation in the difficulty of questions about data visualizations,'' \emph{arXiv preprint arXiv:2505.08031}, 2025.

\bibitem[Chang et~al.(2025)Chang, Wang, Wang, Zhou, Bulling, and Bearfield]{chang2025tell}
M.~Chang, Y.~Wang, H.~W. Wang, Y.~Zhou, A.~Bulling, and C.~X. Bearfield, ``Tell me without telling me: Two-way prediction of visualization literacy and visual attention,'' \emph{arXiv preprint arXiv:2508.03713}, 2025.

\bibitem[Verma et~al.(2024)Verma, Mukherjee, Potts, Kreiss, and Fan]{verma2024evaluating}
A.~Verma, K.~Mukherjee, C.~Potts, E.~Kreiss, and J.~E. Fan, ``Evaluating human and machine understanding of data visualizations,'' in \emph{Proceedings of the annual meeting of the cognitive science society}, vol.~46, 2024.

\bibitem[Halkiopoulos and Gkintoni(2024)]{halkiopoulosGkintoni2024}
C.~Halkiopoulos and E.~Gkintoni, ``Leveraging ai in e-learning: Personalized learning and adaptive assessment through cognitive neuropsychology—a systematic analysis,'' \emph{Electronics}, vol.~13, no.~18, p. 3762, 2024.

\bibitem[Powers(2007)]{roc}
D.~M.~W. Powers, ``Evaluation: From precision, recall and {F}-measure to {ROC}, informedness, markedness \& correlation,'' \emph{Journal of Machine Learning Technology}, vol.~2, no.~1, pp. 37--63, 2007.

\bibitem[Lessmann et~al.(2008)Lessmann, Baesens, Mues, and Pietsch]{DBLP:journals/tse/LessmannBMP08}
\BIBentryALTinterwordspacing
S.~Lessmann, B.~Baesens, C.~Mues, and S.~Pietsch, ``Benchmarking classification models for software defect prediction: {A} proposed framework and novel findings,'' \emph{{IEEE} Trans. Software Eng.}, vol.~34, no.~4, pp. 485--496, 2008. [Online]. Available: \url{https://doi.org/10.1109/TSE.2008.35}
\BIBentrySTDinterwordspacing

\bibitem[Cohen(1960)]{Cohen1960Kappa}
J.~Cohen, ``A coefficient of agreement for nominal scales,'' \emph{Educational and Psychological Measurement}, vol.~20, no.~1, pp. 37--46, 1960.

\bibitem[Locoro et~al.(2025)Locoro, Golia, and Falessi]{LocoroGoliaFalessi2025}
A.~Locoro, S.~Golia, and D.~Falessi, ``Drive-t: A methodology for discriminative and representative data viz item selection for literacy construct and assessment,'' \emph{arXiv preprint arXiv:2508.04160}, 2025.

\bibitem[Rasch(1960)]{Rasch1960}
G.~Rasch, \emph{Probabilistic models for some intelligence and attainment tests}.\hskip 1em plus 0.5em minus 0.4em\relax Danish Institute for Educational Research. (Expanded edition, 1980. Chicago: University of Chicago Press.), 1960.

\bibitem[Wright and Masters(1982)]{WrightMasters1982}
B.~D. Wright and G.~N. Masters, \emph{Rating scale analysis}.\hskip 1em plus 0.5em minus 0.4em\relax Chicago: MESA PRESS, 1982.

\bibitem[Bond and Fox(2015)]{BondFox2015}
T.~G. Bond and C.~M. Fox, \emph{Applying the Rasch Model: Fundamental Measurement in the Human Sciences, Third Edition}.\hskip 1em plus 0.5em minus 0.4em\relax New York: Routledge, 2015.

\bibitem[Linacre(2023)]{WINSTEP}
\BIBentryALTinterwordspacing
J.~M. Linacre, ``Winsteps\textsuperscript{\textregistered} rasch measurement computer program (version 5.4.0),'' 2023. [Online]. Available: \url{https://www.winsteps.com}
\BIBentrySTDinterwordspacing

\bibitem[Jr. et~al.(2013)Jr., Lemeshow, and Sturdivant]{HosmerLemeshowSturdivant2013}
D.~W.~H. Jr., S.~Lemeshow, and R.~X. Sturdivant, \emph{Applied Logistic Regression. Third Edition}.\hskip 1em plus 0.5em minus 0.4em\relax Hoboken, New Jersey: John Wiley Sons, 2013.

\bibitem[Bishop(2007)]{DBLP:books/lib/Bishop07}
\BIBentryALTinterwordspacing
C.~M. Bishop, \emph{Pattern recognition and machine learning, 5th Edition}, ser. Information science and statistics.\hskip 1em plus 0.5em minus 0.4em\relax Springer, 2007. [Online]. Available: \url{https://www.worldcat.org/oclc/71008143}
\BIBentrySTDinterwordspacing

\bibitem[Breiman(2001)]{DBLP:journals/ml/Breiman01}
\BIBentryALTinterwordspacing
L.~Breiman, ``Random forests,'' \emph{Mach. Learn.}, vol.~45, no.~1, pp. 5--32, 2001. [Online]. Available: \url{https://doi.org/10.1023/A:1010933404324}
\BIBentrySTDinterwordspacing

\bibitem[Hall(2000)]{DBLP:conf/icml/Hall00}
M.~A. Hall, ``Correlation-based feature selection for discrete and numeric class machine learning,'' in \emph{ICML 2000}, 2000, pp. 359--366.

\bibitem[Blum and Langley(1997)]{DBLP:journals/ai/BlumL97}
\BIBentryALTinterwordspacing
A.~Blum and P.~Langley, ``Selection of relevant features and examples in machine learning,'' \emph{Artif. Intell.}, vol.~97, no. 1-2, pp. 245--271, 1997. [Online]. Available: \url{https://doi.org/10.1016/S0004-3702(97)00063-5}
\BIBentrySTDinterwordspacing

\bibitem[Guyon and Elisseeff(2003)]{DBLP:journals/jmlr/GuyonE03}
I.~Guyon and A.~Elisseeff, ``An introduction to variable and feature selection,'' \emph{J. Mach. Learn. Res.}, vol.~3, pp. 1157--1182, 2003.

\bibitem[Quinlan(1993)]{DBLP:books/mk/Quinlan93}
J.~R. Quinlan, \emph{C4.5: Programs for Machine Learning}.\hskip 1em plus 0.5em minus 0.4em\relax Morgan Kaufmann, 1993.

\bibitem[Kohavi(1995)]{DBLP:conf/ijcai/Kohavi95}
R.~Kohavi, ``A study of cross-validation and bootstrap for accuracy estimation and model selection,'' in \emph{IJCAI 1995}, 1995, pp. 1137--1145.

\bibitem[Varma and Simon(2006)]{DBLP:journals/bmcbi/VarmaS06}
S.~Varma and R.~Simon, ``Bias in error estimation when using cross-validation for model selection,'' \emph{BMC Bioinform.}, vol.~7, p.~91, 2006.

\bibitem[Cawley and Talbot(2010)]{DBLP:journals/jmlr/CawleyT10}
\BIBentryALTinterwordspacing
G.~C. Cawley and N.~L.~C. Talbot, ``On over-fitting in model selection and subsequent selection bias in performance evaluation,'' \emph{J. Mach. Learn. Res.}, vol.~11, pp. 2079--2107, 2010. [Online]. Available: \url{https://doi.org/10.5555/1756006.1859921}
\BIBentrySTDinterwordspacing

\bibitem[Frank et~al.(2005)Frank, Hall, Holmes, Kirkby, and Pfahringer]{DBLP:books/sp/datamining2005/FrankHHKP05}
E.~Frank, M.~A. Hall, G.~Holmes, R.~Kirkby, and B.~Pfahringer, ``Weka - a machine learning workbench for data mining,'' in \emph{The Data Mining and Knowledge Discovery Handbook}, 2005, pp. 1305--1314.

\bibitem[Witten et~al.(2011)Witten, Frank, and Hall]{DBLP:books/lib/WittenFH11}
I.~H. Witten, E.~Frank, and M.~A. Hall, \emph{Data mining: practical machine learning tools and techniques, 3rd Edition}.\hskip 1em plus 0.5em minus 0.4em\relax Morgan Kaufmann, Elsevier, 2011.

\bibitem[Falessi et~al.(2023)Falessi, Laureani, {\c{C}}arka, Esposito, and da~Costa]{DBLP:journals/ese/FalessiLCEC23}
\BIBentryALTinterwordspacing
D.~Falessi, S.~M. Laureani, J.~{\c{C}}arka, M.~Esposito, and D.~A. da~Costa, ``Enhancing the defectiveness prediction of methods and classes via {JIT},'' \emph{Empir. Softw. Eng.}, vol.~28, no.~2, p.~37, 2023. [Online]. Available: \url{https://doi.org/10.1007/s10664-022-10261-z}
\BIBentrySTDinterwordspacing

\bibitem[Falessi et~al.(2020)Falessi, Roll, Guo, and Cleland{-}Huang]{DBLP:journals/tse/FalessiRGC20}
\BIBentryALTinterwordspacing
D.~Falessi, J.~Roll, J.~L.~C. Guo, and J.~Cleland{-}Huang, ``Leveraging historical associations between requirements and source code to identify impacted classes,'' \emph{{IEEE} Trans. Software Eng.}, vol.~46, no.~4, pp. 420--441, 2020. [Online]. Available: \url{https://doi.org/10.1109/TSE.2018.2861735}
\BIBentrySTDinterwordspacing

\bibitem[Ge et~al.(2024)Ge, Hedayati, Cui, Ding, Bonilla, Joshi, Ottley, Bach, Kwon, Rapp, et~al.]{Ge2024Visualization}
L.~W. Ge, M.~Hedayati, Y.~Cui, Y.~Ding, K.~Bonilla, A.~Joshi, A.~Ottley, B.~Bach, B.~C. Kwon, D.~N. Rapp \emph{et~al.}, ``Toward a more comprehensive understanding of visualization literacy,'' in \emph{Proceedings of the CHI Conference on Human Factors in Computing Systems}.\hskip 1em plus 0.5em minus 0.4em\relax New York, NY, USA: ACM, May 2024, p.~7.

\bibitem[Lily et~al.(2025)Lily, Cabouat, Bonilla, Cui, Ding, Rakotondravony, Creamer, Otto, Hedayati, Kwon, et~al.]{lily2025autoethnography}
W.~G. Lily, A.-F. Cabouat, K.~Bonilla, Y.~Cui, Y.~Ding, N.~Rakotondravony, M.~M. Creamer, J.~T. Otto, M.~Hedayati, B.~C. Kwon \emph{et~al.}, ``An autoethnography on visualization literacy: A wicked measurement problem,'' \emph{IEEE Transactions on Visualization and Computer Graphics}, 2025.

\bibitem[Kempt(2024)]{kempt2024explainable}
H.~Kempt, \emph{(Un) explainable Technology}.\hskip 1em plus 0.5em minus 0.4em\relax Springer Nature, 2024.

\end{thebibliography}
\end{document}